\documentclass[12pt,prepint]{aastex}
\usepackage{amsmath}
\usepackage{natbib}
\usepackage{multirow}
\usepackage{colortbl}
\usepackage{apjfonts}

\begin{document}

\shorttitle{}
\shortauthors{Sun et al.}

\title{The Discovery of Timescale-Dependent Color Variability of Quasars}

\author{Yu-Han Sun\altaffilmark{1}, Jun-Xian Wang\altaffilmark{1}, Xiao-Yang Chen\altaffilmark{1} and Zhen-Ya Zheng\altaffilmark{2}}

\altaffiltext{1}{CAS Key Laboratory for Research in Galaxies and Cosmology, Department of Astronomy,\\ University of Science and Technology of China, Hefei, Anhui 230026, China. sunyh92@mail.ustc.edu.cn; jxw@ustc.edu.cn}
\altaffiltext{2}{School of Earth and Space Exploration, Arizona State University, Tempe, AZ 85287}
\email{sunyh92@mail.ustc.edu.cn; jxw@ustc.edu.cn}


\begin{abstract}
Quasars are variable on timescales from days to years in UV/optical, and generally appear bluer while they brighten. The physics behind the variations in fluxes and colors remains unclear.
Using SDSS $g$ and $r$ band photometric monitoring data for quasars in Stripe 82, we find that although the flux variation amplitude increases with timescale, the color variability exhibits opposite behavior.
The color variability  of quasars is prominent at timescales as short as $\sim$ 10 days, but gradually reduces toward timescales up to years. 
In other words, the variable emission at shorter timescales is bluer than that at longer timescales.
This timescale dependence is clearly and consistently detected at all redshifts from z = 0 to 3.5, thus can not be due to contaminations to broadband photometry from emission lines which do not respond to fast continuum variations.
The discovery directly rules out the possibility that simply attributes the color variability to contamination from a non-variable redder component, such as the host galaxy.
It can not be interpreted as changes in global accretion rate either.
The thermal accretion disk fluctuation model is favored, in the sense that fluctuations in the inner hotter region of the disk are responsible for short term variations,
while longer term and stronger variations are expected from the larger and cooler disk region.
An interesting implication is that one can use quasar variations at different timescales to probe disk emission at different radii.
\end{abstract}

\section{Introduction}
Active galactic nuclei (AGNs), including quasars and their low luminosity analogs, have been found to be variable from radio to X-ray and Gamma ray. It is widely accepted that the optical and UV emission of quasars, which are variable on timescales of days to years, are dominated by thermal emission (i.e. the big blue bump) from the accretion disk surrounding the central supper massive black hole. Note there are still challenges from observations to accretion disk models \citep[e.g.][]{Davis2007,Bonning2007,Lawrence2012}.

Studying the variations of UV/optical emission provides a useful approach to probe the physics of the accretion disk. Meanwhile, variations are also essential to reverberation mapping studies, including the mapping of the broad emission line region, dusty torus, and accretion disk \citep[e.g.][]{Peterson1993,Suganuma2006,Collier1999}. AGNs can also be efficiently selected from photometric surveys based on variations \citep[e.g.][]{Sesar2007}.

The UV/optical variations in AGNs are generally stochastic and could be
modeled with a damped random walk process \citep{Kelly2009,Kozlowski2010,Macleod2010,Zu2013} on timescales of weeks to years.
Also note that on shorter timescales, the power density spectra of Kepler AGNs appear much steeper with slopes of -2.6 to -3.3 \citep{Mushotzky2011}, comparing to brown noise ($S$ $\sim$ $f^{-2}$) from random walk processes.

The UV/optical variations in AGNs are also wavelength dependent. AGNs generally show larger variation amplitudes in bluer bands \citep{Cristiani1997,Giveon1999,Hawkins2003,Berk2004}. The wavelength dependent variation pattern is also seen via a different concept that AGNs normally appear bluer when they get brighter \citep{Giveon1999,Wilhite2005,Sakata2011,Schmidt2012,Wamsteker1990,Webb2000}, showing the variable emission is bluer than the stable or less variable ones.

The nature and origin of such a ``bluer-when-brighter'' pattern is still under debate.
Contaminations from the host galaxies may produce a ``bluer-when-brighter'' pattern as the host galaxy emission is redder and stable \citep[e.g.][]{Hawkins2003}.
Pollution to spectra or broadband photometry by less variable components, such as the small blue bump \citep{Paltani1996} and emission lines may also produce a ``bluer-when-brighter'' pattern, but only in certain wavelength ranges, and may even produce the reverse pattern (``redder-when-brighter'') at different wavelength ranges \citep[e.g.][]{Schmidt2012}.
Besides, the spectra of the variable emission in AGNs
appear consistent with predictions from accretion disk theory \citep{Pereyra2006,Li2008,Sakata2011,Zuo2012,Gu2013}. This stimulates assumptions which attribute AGN variations to changes in mass accretion rates.

In this work, using SDSS monitoring data of quasars in Stripe 82, we find that the color variation pattern in quasars is timescale dependent, providing a new and unique vision to the understanding of the color variations in AGNs.
We present our analysis approach in \S2, and the results in \S3. In \S4 we discuss the implications of this interesting discovery.

\section{Methodology and Data}

A ``bluer when brighter'' tendency in the UV/optical color variations in quasars has been well established by linear fitting multi-epoch measurements in magnitude-color space \citep[e.g.][]{Giveon1999,Berk2004,Wilhite2005}. The trend has been confirmed by fitting the color variability in magnitude-magnitude (or flux-flux) space, which can remedy biases due to co-variances between the color and the magnitude uncertainties \citep{Schmidt2012}.

\begin{figure}
\centering
\plotone{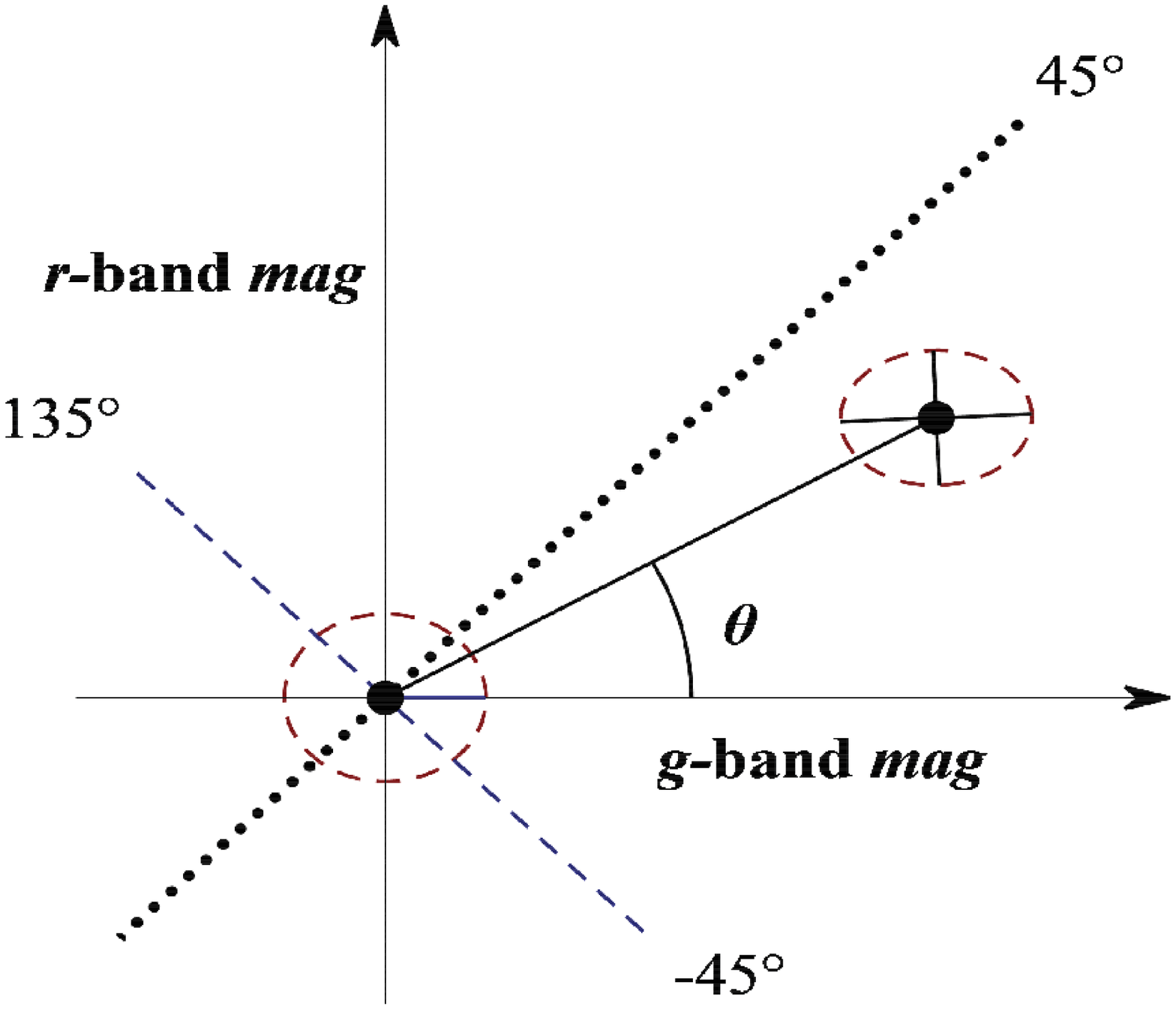}
\caption{Sketch of $\theta$ in $g$-$r$ (magnitude-magnitude) space. The arrows of the x and y axes point to brighter fluxes. Two solid black dots demonstrate two photometry measurements of a single source in the $g$ and $r$ bands, and the ellipses surrounding the the black dots stand for the photometric uncertainties in the $g$ and $r$ bands. We use the the inclination $\theta$ of the solid line connecting two black dots to quantify the color variation between two epochs, with 45$^{\circ}$ $>$ $\theta$ $\geq$ 0$^{\circ}$ standing ranges for ``bluer when brighter'', and 90$^{\circ}$ $\geq$ $\theta$ $>$ 45$^{\circ}$ for ``redder when brighter''. The dotted line with an inclination of 45$^{\circ}$ plots variations with no color changes.
Due to uncertainties in photometry measurements, $\theta$ also likely falls in the ranges of -45$^{\circ}$ to 0$^{\circ}$ (``bluer when brighter"), and 90$^{\circ}$ to 135$^{\circ}$ (``redder when brighter").
The range of $\theta$ is limited to [-45$^{\circ}$,135$^{\circ}$], within which, the smaller the $\theta$ is, the stronger
the ``bluer when brighter'' trend is.
}
\label{theta}
\end{figure}

In order to explore the timescale dependence  of the  color variation in quasars, we develop a structure function approach. For each pair epochs of an individual quasar, we define a parameter $\theta$ (in the mag-mag space, see Fig. \ref{theta}, we use the $g$ and $r$ bands for instance) to measure the amplitude of the color variation between two epochs, and the time lag $\tau$ between two epochs to stand for the timescale of the color variation.
\begin{equation}
\theta(\tau)=\arctan \left(\frac{m_r(t+\tau)-m_r(t)}{m_g(t+\tau)-m_g(t)}\right)
\end{equation}
Quasars usually brighten or darken simultaneously in the $g$ and $r$ bands, thus $\theta$ should fall in the range of [0$^{\circ}$,90$^{\circ}$].  However, there are cases where quasars brighten in the $g$ band but darken in the $r$ band or vise versa (most likely due to photometric uncertainties), yielding $\theta$ in the ranges of [-45$^{\circ}$,0$^{\circ}$] and [90$^{\circ}$,135$^{\circ}$].
We restrict the valid range of $\theta$ to  [-45$^{\circ}$,135$^{\circ}$], and $\theta$ outside of this range
(e.g. [135$^{\circ}$,315$^{\circ}$]) can be
transformed into it by minus 180$^{\circ}$ (see Fig. \ref{theta} and its caption).
Averaging $\theta(\tau)$ over all epoch pairs within given timescale ranges (which can be done for a single quasar with well sampled two bands light curves, or for multiple quasars) we get
\begin{equation}
\bar\theta(\tau)=\frac{\sum\limits_i^N \theta_i(\tau)}{N}
\end{equation}
where N stands for the number of pairs with $\tau$ falling in given ranges.

Note the inclination $\theta$ quantifies the color variability similarly to the slope of the variation in mag-mag space \citep[effectively $\Delta m_r/\Delta m_g$, e.g. as adopted by][]{Schmidt2012}, but in a linear manner (see Fig. \ref{theta}), enabling us to average $\theta$ over a wide range due to  photometric uncertainties.
For example, $\theta$ = 0$^{\circ}$ (90$^{\circ}$) represents variation in the $g$ ($r$) band only, an extreme case of ``bluer (redder) when brighter'' (see Fig \ref{theta}), and corresponds to a slope of the variation
($\Delta m_r/\Delta m_g$) in the mag-mag space of 0 ($\infty$).
Averaging 0$^{\circ}$ and 90$^{\circ}$  yields 45$^{\circ}$, indicating no color variation on average, which is logically reasonable, but averaging the slopes 0 and $\infty$ we still get $\infty$.
This is particularly important for this study since we are measuring color variation from each pair of epochs which produces considerably large scatter in $\theta$ due to photometric uncertainties.
Contrarily, \citet{Schmidt2012} measured the color variability slope using $\sim$ 60 epochs for each quasar, which suffers much less from photometric uncertainties; however, any timescale dependence of the color variation is lost during this approach.

\begin{figure}
\centering
\plotone{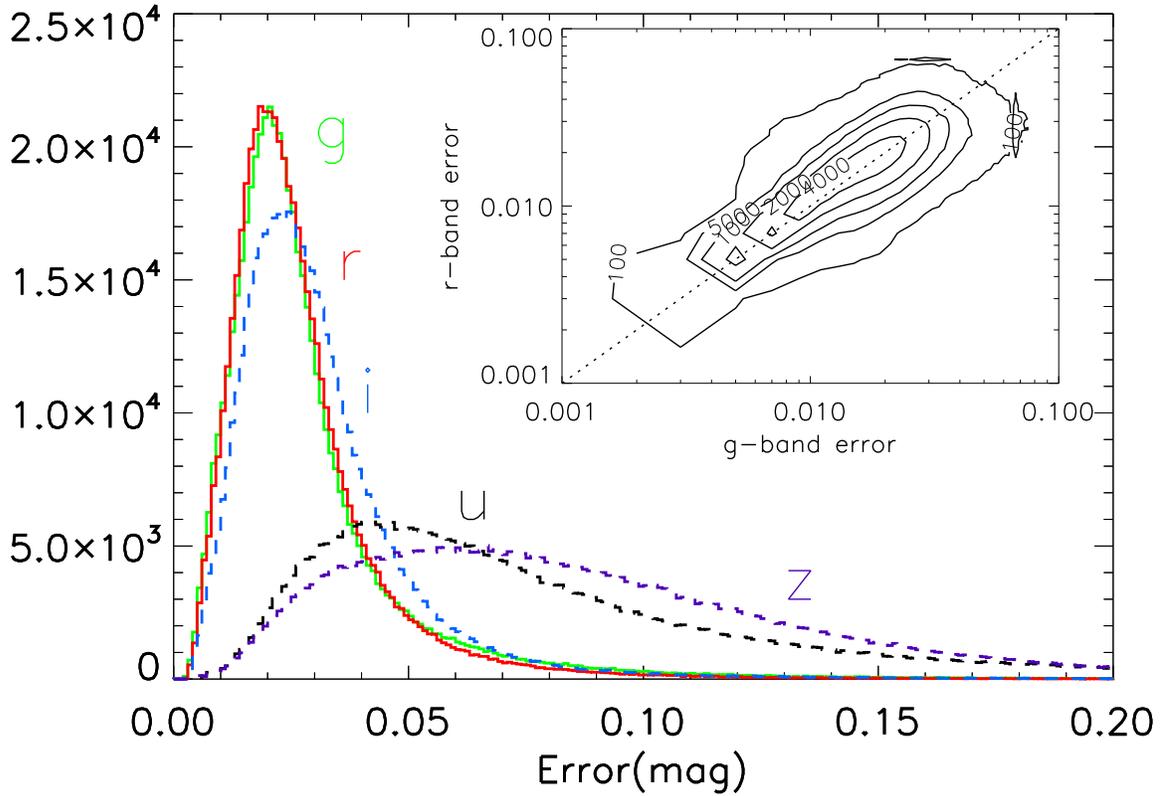}
\caption{Photometric error histogram and contour in the $g$ $\&$ $r$ bands. The range of error in both bands in the contour plot is [0.001,0.1]. It's clear that the overall distributions of the uncertainties in the $g$ and $r$ bands are almost the same (left panel), and most epochs have similar uncertainties in both bands (right panel).  }
\label{magerr}
\end{figure}

We use the data of the Sloan Digital Sky Survey's (SDSS's) Stripe 82 (S82) presented by \citet{Macleod2012}.
Their Southern Sample catalog provides re-calibrated $\sim$ 10 year-long light curves in 5 SDSS bands ($ugriz$) for 9,258 spectroscopically confirmed quasars in DR7.
In this work we focus on the $g$ and $r$ bands in which SDSS have best and comparable photometries for those spectroscopically confirmed quasars
(see Fig. \ref{magerr}).
We reject epochs which have unphysical $g$ or $r$ band magnitudes. We further exclude data points with photometric uncertainties in the $g$ or the $r$ band greater than 0.1 mag, which often appear as outliers in the light curves, and could have been due to bad observation conditions. However, keeping them would not alter any scientific results presented in this work except for slightly increasing  the underlying noise level of the structure function (see Fig. \ref{CV}, upper panel, at timescale less than $\sim$ 30 days).
We select sources with more than 10 good epochs, leading to a final sample of 8,944 quasars.

\section{Results}

\subsection{Timescale Dependent Color Variation}

\begin{figure}
\centering
\plotone{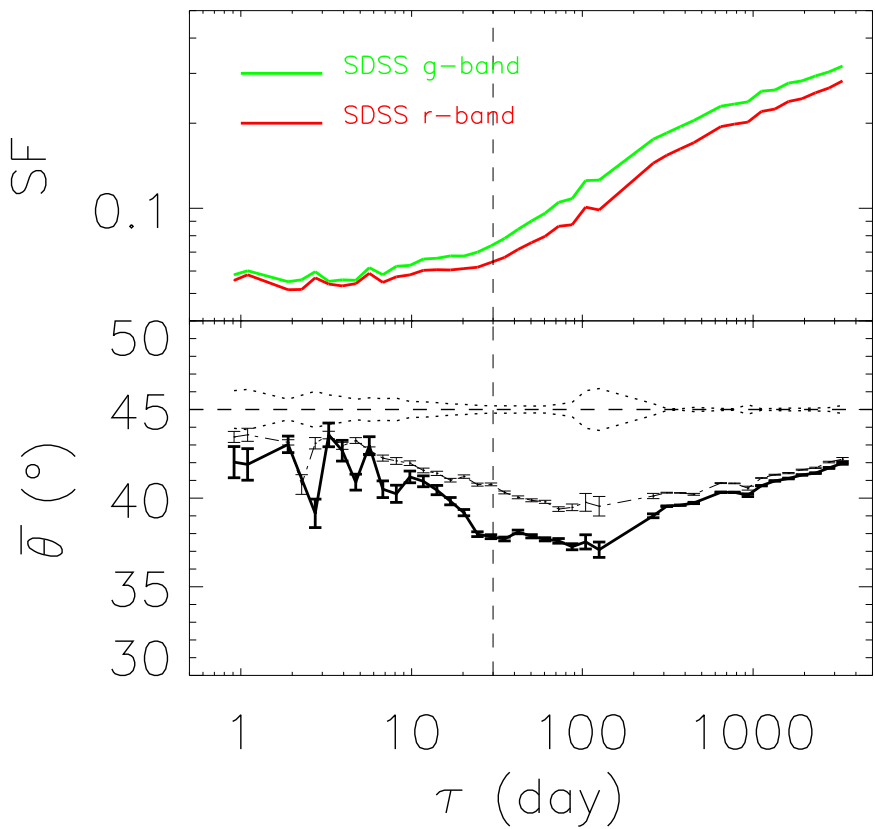}
 \caption{Upper panel: SDSS $g$ and $r$ band ensemble structure functions (in the observed frame) of our sample of quasars in Stripe 82.
Lower panel: The average amplitude of their color variation ($\bar\theta$, see \S2 and Fig. \ref{theta} for definition)
as a function of timescale in the observed frame.
The bin size of $\tau$ is 0.08 in log space.
In both panels, we dropped a couple of data bins (for instance at $\sim$ 200 days) due to too few data pairs available in those timescale bins. In the lower panel, the dashed line at $\bar\theta$ = 45$^{\circ}$ draws the boundary between ``bluer when brighter'' ($<$ 45$^{\circ}$) and ``redder when brighter'' ($>$ 45$^{\circ}$).
The dotted lines plot the standard 1$\sigma$  errors of $\bar\theta$ assuming $\theta$ measured from each data pair distributes randomly within [-45$^{\circ}$,135$^{\circ}$]. Such errors are solely determined by the
number of data pairs in each timescale bin, and can be interpreted as the conservative upper limits to the uncertainties
of observed $\bar\theta$. The observed $\bar\theta$ and its 1$\sigma$ errors are plotted
as the solid and dash-dotted line (after and before applying the ``3$\sigma$'' criterion on the variation respectively).
 }\label{CV}
\end{figure}

In the upper panel of Fig. \ref{CV} we first plot the ensemble  structure function  (without subtracting the photometric uncertainties) of our sample in the $g$ and $r$ bands in the observed frame.
We collect all pair epochs within each light curve of the quasars and calculate the ensemble structure function as
\begin{equation}
		SF(\tau)=sqrt(\frac{1}{N(\tau)}\sum_t{[m(t+\tau)-m(t)]^2)}
    \end{equation}
We clearly see stronger variation in the $g$ band than the $r$ band, indicating the well known pattern of larger variation in bluer band. The amplitudes of the variation in both bands drop with decreasing timescale, and tend to flat at timescale $<$ 30 days. This shows that at smaller timescale the structure function is dominated by photometric uncertainties.
By averaging the structure function below 5 days, we obtain a mean value of 0.057 mag and 0.054 mag for the $g$ and the $r$ band respectively, reflecting the level of photometric accuracy in both bands.

Lower panel in Fig. 3 plots the mean color variability of all 8,944 QSOs as a function of timescale in the observed frame (grey data points and dash-dotted line).
It's clear that there is a significant timescale dependence in the color variability of QSOs:
while all the data points lie below the $\theta$ = 45$^{\circ}$ line, suggesting the quasars get bluer when they brighten at different timescales, the ``bluer when brighter'' trend gradually weakens with  increasing timescale above 100 days.
We note that $\bar\theta$ also increases below 30 days, which is likely due to contamination by the photometric uncertainties, as random errors alone in the $g$ and $r$ bands (note $g$ and $r$ band have comparable photometries, see Fig. \ref{magerr}) would yield $\bar\theta$ = 45$^{\circ}$.

To minimize the pollution from photometric uncertainties, we further exclude pair epochs in which the flux variation in the $g$ - $r$ space is statistically insignificant ($<$ 3 $\sigma$). The 1 $\sigma$ uncertainty of the variation between two epochs in the mag-mag space was calculated by treating the photometric uncertainties of each epoch in the $g$ - $r$ space as ellipses, and adding the uncertainties within two ellipses along the variation quadratically (see Fig. \ref{theta}).
The fraction of the excluded pair epochs with this approach decreases gradually from $\sim$ 90\% at timescale of 1 -- 10 days, to $\sim$ 20\% at timescale above 1000 days. Note this criterion is adopted to exclude pair epochs with insignificant flux variations, independent to the significance of the color variation.
The new output $\bar\theta$ is significantly smaller at intermediate timescales (thick solid line in Fig. \ref{CV}), confirming that photometric uncertainties could smear out the color
variation at small timescales. Instead, at timescales $<$ 30 days, quasars likely have even stronger ``bluer when brighter"
tendency, however, due to the much weaker variation in fluxes at shorter timescales, this can only be examined with data points with much better photometry, such as with space-borne observations.
We also note that the 3$\sigma$ cut is biasing the selection to stronger flux variations at short timescales.
Assuming at given timescales the flux variations with different amplitudes behave similarly in color variability
(i.e., with the same $\theta$), this bias would not affect the study in this work. Actually,  \citet{Schmidt2012}
have shown that at least at long timescales, lower-amplitude flux variations tend to produce more color variability.
If the same trend holds at short timescales, we should expect intrinsically even stronger short term color variability, further strengthening the results of this work.
Hereafter we keep the approach of excluding epochs with flux variation $<$ 3 $\sigma$ in mag-mag space and limit our scope to timescales $>$ 30 days (in the observed frame).

\subsection{Redshift Dependence}
The rest frame wavelength ranges probed with SDSS $g$ and $r$ bands change substantially with redshift. Furthermore, it is well known that the emission lines (and Balmer continuum) in the spectra, which do not respond to fast central continuum variations, contribute differently to the broad band photometry at various redshifts \citep[e.g.][]{Wilhite2005,Schmidt2012}.
Figure 4 plots the color variability $\bar\theta$ for the 8,944 spectroscopically confirmed S82 QSOs in different observed timescale ranges as a function of redshift. The color variation shows a redshift dependent pattern similar to Fig. 4 of \cite{Schmidt2012}, who has been nicely attributed to contribution from the emission line components in the quasar spectra which do not respond to continuum variation. For instance,  the distributions of $\bar\theta$ peak at z $\sim$ 0.7 where MgII line is in the $g$ band; and drop at z $>$ 0.95 because MgII line moves from the $g$ to the $r$ band, and reach the minimum  point at z $\sim$ 1.4 while MgII line moves out from the $r$ band.
However, we see that the color variation is clearly stronger at shorter timescale at all given redshifts. This proves that the timescale dependent color variation pattern we showed in Fig. \ref{CV} is not an artificial effect related to redshift.

To better demonstrate the timescale dependence of the color variation at different redshifts, we divide the quasars into different redshift bins (selected by putting H$\beta$, MgII, CIII] or Ly$\alpha$ lines into the $g$ or $r$ band accordingly).
In Fig. \ref{zbin} we clearly see similar timescale dependence of the color variation in different redshift bins.
This further confirms that the timescale dependence reflects the variation pattern of the continuum, but can not be attributed to emission lines in the spectra which only respond to variations at longer timescales.

\section{Discussion}

We have shown with SDSS Stripe 82 monitoring data that the color variation (the ``bluer when brighter'' trend) in quasars is more prominent at short timescales of $\sim$ 10 days, and gradually reduces toward longer timescales up to several years.
The ``bluer when brighter'' pattern may even disappear at much longer timescales if we simply extend the lines in \ref{CV}.
This directly indicates that shorter term variations in quasars are significantly bluer, although with smaller amplitudes in flux variation, than longer term ones.
This is because the color of the varied emission
\begin{equation}
\frac{\Delta f_r}{\Delta f_g} = \frac{10^{0.4(m_r-\Delta m_r)}-10^{0.4m_r}}{10^{0.4(m_g-\Delta m_g)}-10^{0.4m_g}}
\approx \frac{f_r}{f_g} \times \frac{\Delta m_r}{\Delta m_g} = \frac{f_r}{f_g} \times \tan \theta(\tau)
\end{equation}

The immediate subsequences of this interesting discovery and its possible physical nature are discussed below.

\subsection{Can we simply attribute the ``bluer when brighter" pattern to contamination from the host galaxies?}

Mixing a variable component in bluer but constant color with a stable redder spectrum, such as from the host galaxy, could naturally produces a ``bluer when brighter'' pattern in AGNs \citep[e.g.][]{Hawkins2003}.
Furthermore, if this is the only or dominated mechanism behind the color variation in AGNs, one can utilize a flux variation gradient method to distinguish the intrinsic spectrum of the disk (variable) emission from the (non-variable) host galaxy emission \citep[e.g.][]{Choloniewski1981,Winkler1992,Winkler1997,Fozo2013}.
This possibility has been studied and discussed by many works, and contrary results have been reported \citep[e.g.][]{Woo2007,Walsh2009,Sakata2010,Sakata2011}.  For instance,  \citet{Sakata2011} found that contributions from the host galaxies are too weak to explain the observed ``bluer-when-brighter'' pattern in a small sample of 10 quasars, while \citet{Sakata2010} showed that host galaxy plus narrow emission lines are sufficient to explain the long term ``bluer when brighter'' pattern in 11 nearby AGNs.
Generally, the relative importance of the contamination from the host galaxy
is wavelength and AGN type dependent, which may help to partly explain the inconsistencies between different studies.

However, our discovery of the timescale-dependent color variations could immediately rule out the possibility that simply attributing the ``bluer-when-brighter'' pattern in AGNs to mixture of a variable disk emission with blue but constant color and a redder stable emission such as from the host galaxy, since this model predicts
that the color of the variable emission is timescale independent. Furthermore, the approach to use the flux variation gradient method to separate AGN disk emission from the host galaxy contribution should be rigorously invalid.

Naturally we shall expect the color variations of Seyfert galaxies, the less luminous analog of quasars, should be similarly timescale dependent. If so, this can help to solve the discrepancies in observations which show that the host galaxy emission is insufficient to
explain the observed color variations in quasars \citep[e.g.][]{Sakata2011}, while the long term color variations  in Seyfert galaxies are rather weak or non-detectable after subtracting host galaxy contribution \citep[e.g.][]{Woo2007,Walsh2009,Sakata2010}. We have shown that in SDSS quasars the color variations are prominent at timescale of $\sim$ 30 days, but much weaker at timescale of a couple of years. For Seyfert galaxies which harbor less massive black holes comparing with quasars, the corresponding timescales could be lower by a factor $\sim$ 100 (black hole mass of 10$^7$ M$_\odot$ versus 10$^9$ M$_\odot$ for instance).
Hence the color variations in Seyfert galaxies are expected to be much weaker at given observed timescales comparing with quasars.

\subsection{Changes in Global Accretion Rate?}

\citet{Pereyra2006} reported that the variable quasar spectrum (the composite differential spectrum between two epochs of observations for hundreds of SDSS quasars) in UV/optical could be well fitted by a standard accretion disk model with changes in accretion rate. They therefore proposed that the variations in quasars could be simply caused by changes in global accretion rates and quasars appear bluer when they brighten because of higher disk temperature in case of larger accretion rate. This scenario is further supported by later studies \citep{Li2008,Sakata2011,Zuo2012,Gu2013}. Note this interpretation relies on the validity of the assumption that time scale in global accretion rate variation could be as short as days to months, which is under debates since the viscous timescale of the accretion disk is expected to be much longer in AGNs.

In this scenario, the color and flux of each quasar are simply determined by its global accretion rate,
and its variation follows a fixed smooth monotonic curve in $g$--$r$ mag-mag space.
The local slope of the curve represents the color variability $\theta$ as shown in Fig. \ref{theta},
and in this scheme short term variations simply accumulate into long term ones along the fixed curve in mag-mag space.
If short term color variability is strong  (i.e. with small $\theta$), the accumulated long term color variability should have similarly small $\theta$. In other words, the color variability behavior is expected to be timescale independent.
This is directly rejected by the finding of this work, indicating the observed variation in quasars can not be simply explained as changes in accretion rate. This conclusion is also consistent with a few recent works which either show that accretion rate changes
can not model the the variation emission of quasars in Stripe 82 \citep{Ruan2014,Kokubo2014}, or the color variations in individual quasars are much more pronounced than the range of color seen in quasars with different accretion rates \citep{Schmidt2012}.

\subsection{The Nature of the Variation and Color Variability}

The fact that long term optical light curves of quasars could be described as damped random walk processes
with characteristic timescales comparable with thermal timescales, suggests the variations are
due to thermal fluctuations in the accretion disk, likely driven by magnetic field related stochastic process \citep[e.g.][]{Kelly2009,Macleod2010}.
Interestingly, \citet{Ruan2014} proposed that the spectral (color) variability of quasars could also be reproduced by a simple  inhomogeneous disk model with large localized temperature fluctuations.

Within this frame, we discuss implications of the discovery presented in this work.
As we have pointed out, the timescale dependence of color variations indicates long term variations can not be simply regarded as
accumulation with time of stochastic short term ones. Instead, variations at different timescales have different colors.
A natural explanation is that short term variations are dominated by thermal fluctuations in the inner most region of the accretions disk where the disk  is hotter and the disk emission is bluer, while longer term variations are produced over larger scales with lower effective disk temperature.
Short term fluctuations can also occur at larger disk radius, but could have been smeared out by large number of random events
over large disk surface; and only longer term variations could be detected at larger physical scale, yielding less bluer
variations.
It is also possible that the disk fluctuations propagate from inner to outer region, and only fluctuations with larger amplitudes could propagate to larger radii and last longer.
This may also explain the variation in fluxes increases with timescale, and could be testified with Monte Carlo simulations.
But note the velocity of the propagation is required to be very fast, even close to the speed of light, required by the simultaneity of multi-color variations \citep[e.g.][]{Gaskell2008}.
Meanwhile, an immediate consequence is that one can use variations at different timescales to probe the disk emission at different radii. The timescale dependence could also be used to constrain/refine the disk fluctuation models \citep[e.g.][]{Dexter2011,Ruan2014}. Further more extensive studies on these issues are under-going, and will be presented in forthcoming paper(s).

\acknowledgments
This work is supported by Chinese NSF (grant No. 11233002) and the Strategic Priority Research Program ``The Emergence of Cosmological Structures" of the Chinese Academy of Sciences (grant No. XDB09000000). J.X.W. acknowledges support from Chinese Top-notch Young Talents Program.

\bibliography{color}

\newpage

\begin{figure}
\centering
\plotone{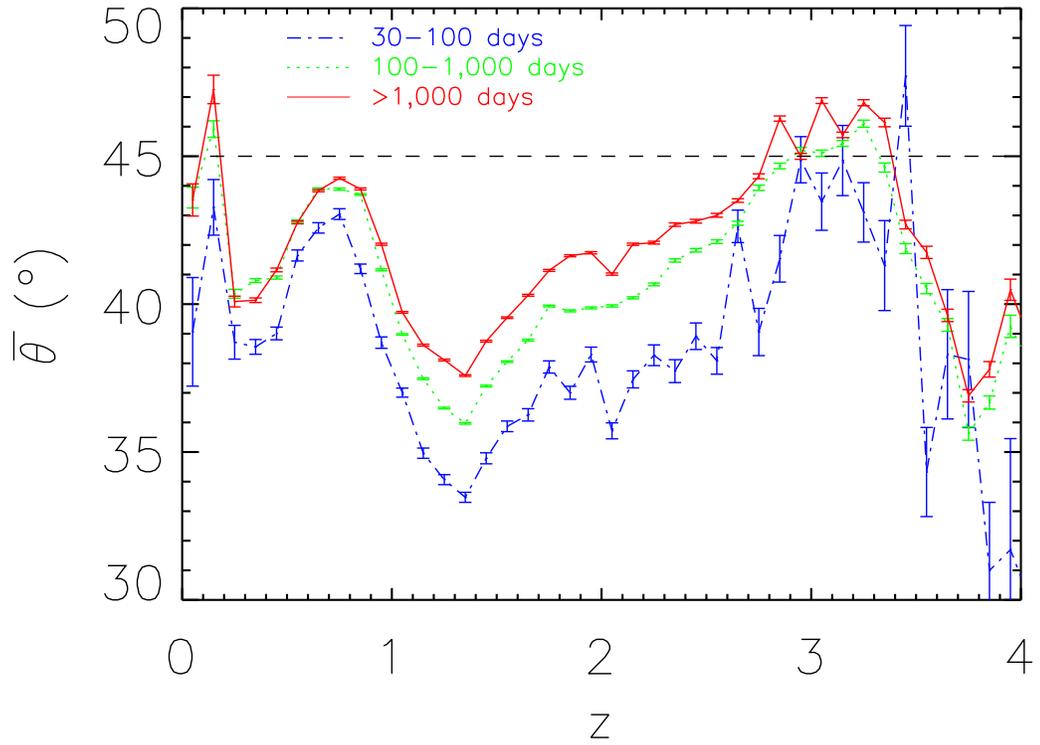}
\caption{The amplitudes of color variation of Stripe 82 quasars (in three different timescale bins in the observed frame)
as a function of redshift. While the color variation is clearly redshift dependent \citep[also see][]{Schmidt2012}, shorter term variation is obviously bluer than longer term one at all given redshifts.
}
\label{redshift}
\end{figure}

\begin{figure}
\centering
\plotone{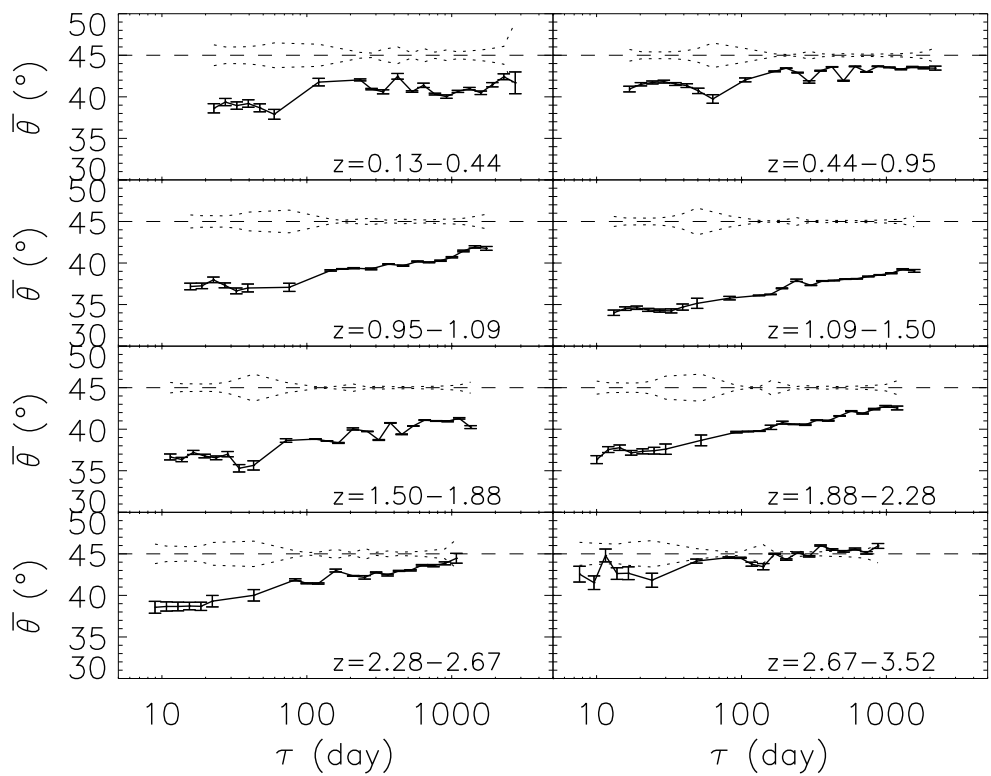}
\caption{The amplitudes of color variation of Stripe 82 quasars as function of timescale in the rest frame in different redshift bins. The symbols are similar to which used in the lower panel of Fig. \ref{CV}, but we drop $\bar\theta$ before applying the ``3$\sigma$'' criterion on the variation. The timescale dependence of the color variation is clearly seen in all redshift bins.
}
\label{zbin}
\end{figure}

\end{document}